\documentclass[%
 reprint,
 superscriptaddress,
 preprintnumbers,
 nofootinbib,
 amsmath,amssymb,
 aps,
 longbibliography,
]{revtex4-2}

\usepackage{graphicx}
\usepackage{dcolumn}
\usepackage{bm}
\usepackage{xcolor}
\usepackage{array}
\usepackage{hyperref}
\usepackage{float}
\usepackage{enumitem}

\usepackage{booktabs}
\usepackage{multirow}
\usepackage{pgfplots}
\usetikzlibrary{shapes.geometric, arrows.meta, positioning}
\pgfplotsset{compat=1.18}

\pgfplotsset{empty legend/.style={legend image code/.code={}}}

\definecolor{llmblue}{RGB}{58,110,165}
\definecolor{llmpink}{RGB}{225,85,136}
\definecolor{llmgreen}{RGB}{91,168,74}
\definecolor{darkyellow}{RGB}{204,153,0}

\newsavebox{\propbox}

\begin{document}

 \preprint{IPMU26-0029}

\title{AI's Capability in Assisting Scientific Research in Physics, Astrophysics, and Cosmology I: Literature Review}
\author{Anamaria Hell$^{1,2}$}
\email{anamaria.hell@ipmu.jp}
\noaffiliation
\author{Kateryna Vovk$^{1,2}$}
\noaffiliation
\author{Veena Krishnaraj$^{3}$}
\noaffiliation
\author{Jia Liu$^{1,2}$}
\noaffiliation
\author{Kosuke Aizawa$^{4,1,2}$}
\noaffiliation
\author{Adrian E.~Bayer$^{5,3}$}
\noaffiliation
\author{Linda Blot$^{1,2}$}
\noaffiliation
\author{Jessica Cowell$^{6,1,2}$}
\noaffiliation
\author{Suyog Garg$^{7}$}
\noaffiliation
\author{Jonathan Gr\'ee$^{8}$}
\noaffiliation
\author{Ben Horowitz$^{1,2}$}
\noaffiliation
\author{Masaya Ichikawa$^{9}$}
\noaffiliation
\author{Kanyuni Iemoto$^{10}$}
\noaffiliation
\author{Keigo Kondo$^{11,7}$}
\noaffiliation
\author{Zacharie Lorsin$^{8}$}
\noaffiliation
\author{Kevin McCarthy$^{1,2}$}
\noaffiliation
\author{Jamie Robinson$^{3}$}
\noaffiliation
\author{Miguel Ruiz-Granda$^{12,13}$}
\noaffiliation
\author{Leander Thiele$^{1,2}$}
\noaffiliation
\author{Ievgen Vovk$^{14}$}
\noaffiliation
\author{Mingshen Zhou$^{15,16}$}
\noaffiliation
\collaboration{Author affiliations are listed at the end of the paper.}
\noaffiliation
\begin{abstract}
We investigate how well large language models (LLMs) can assist with literature reviews for scientific research. We perform a controlled study of eight expert-conceived research projects across the areas of physics, astrophysics, and cosmology. Each project has a defined background and goal, and human experts and AI prompters are asked to perform identical literature review tasks in parallel. We compare the relevant literature selected by humans with that selected by mid-2025 LLMs (ChatGPT-4o, ChatGPT Deep Research, and Gemini). We find the overlap between human- and AI-selected references to be small ($<$6\%), indicating that AI models do not yet reproduce a competent expert search on their own, though they have the potential to complement literature searches by humans. We then assess the reliability and completeness of AI-generated candidate references, distinguishing two types of hallucination: fabrications (references to nonexistent papers) and metadata mismatches (real papers with one or more incorrect fields). We find that while fabricated references make up 3\% of the AI-generated references, 64\% are real papers with at least one incorrect field (title, author, year, journal, DOI, or link), indicating that the mid-2025 models require systematic verification. However, the performance is significantly improved for the 2026 model ChatGPT Pro 5.5, with a single-project test showing zero fabrication or metadata mismatches.
\end{abstract}
\maketitle
\section{Introduction}

The scientific workflow in physics generally consists of four parts: the literature review, project planning, the execution of the study, and writing the paper. Among them, the literature review is especially important. It situates the work among the broader landscape of existing studies, provides means of assessing the originality of the manuscript in preparation, and improves its quality, potentially also giving rise to new ideas. However, in recent years, the number of submitted papers per day has been drastically increasing. The sheer volume of references makes it easier to omit important works. This is especially critical in the early stages of a project, when a researcher new to the field seeks only a rapid assessment of whether the idea in question has already been pursued, without yet committing to an exhaustive analysis of the literature. This leaves considerable room for improvement, and opens the possibility of exploring tools and collaborations that render the search stage more reliable and comprehensive.

Artificial intelligence (AI), and large language models (LLMs) in particular, have undergone rapid progress in just a couple of years. This has immediate consequences for science since AI is increasingly being used for all stages of scientific research. LLMs are being increasingly used in literature reviews \cite{Scherbakov_2025}, have been shown to provide feedback on papers that researchers often find useful \cite{Liang2024}, and have shown promise to perform scientific tasks \cite{2026arXiv260121165W}. They also exhibit weaknesses such as hallucinations, sometimes completely or partially fabricating information (see \cite{11132297} for a review). Such behavior was proven to be highly prevalent in the older generations (2023 -- 2025) of ChatGPT and Gemini models, when asked to retrieve references across different scientific fields \cite{Walters_2023, Franzoni_2024, Pastucha_2025}.

Deep learning has also found significant applications in physics, astrophysics, and cosmology. Starting with numerical applications of AI to accelerate computations and achieve more accurate parameter inferences (e.g., \cite{Hezaveh2017,VillaescusaNavarro2021}), LLMs have also extended AI applications to language- or reasoning-related tasks \cite{Nguyen2023,Zhou2024,Mitchell2023,Gao2023,Lu2024,Shcherbiak2024,Panickssery2024,Si2024,Wataoka2024,Liang2024,Sadasivan2023,Ren2025,Sikimic2025,Sandstrom2026, Thorne2026, denario, Miao:2026gmx,Hell:2026dqx}, suggesting that LLMs can act as cognitive partners in scientific reasoning.
What remains largely unexplored is a direct comparison between human and AI performance in the scientific workflow, set at the frontiers of physics research. In this work, we will study this question with a focus on the literature review, while in the companion paper \cite{Paper2} we will examine AI's capability in project planning and proposal evaluation.

The primary questions that we aim to explore in this work are: Can LLMs help physicists at the frontiers of research to discover literature for a novel scientific project? How many of the AI-generated candidate references are deemed relevant for the study by the expert? How reliable are the references that LLMs provide? This knowledge is especially important for researchers entering a new field, who seek a rapid overview of whether their idea is novel before committing to an exhaustive literature search.

To answer these questions, we conducted a controlled study on eight expert-conceived research projects spanning physics, astrophysics, and cosmology. For each project, the literature search was independently carried out by a human and by LLMs available in mid-2025 -- ChatGPT-4o, ChatGPT Deep Research, and Gemini (2.5-flash, default, which was available in June/July 2025) -- supplemented with project-specific context. The relevance of the AI-generated candidate references was then evaluated by an expert. The overall material was also compared to human output and further analyzed to assess reliability and possible hallucinations. Finally, the literature search was repeated for one of the projects using ChatGPT Pro 5.5 to spot-test the AI's capabilities close to the time of submission.

The remainder of the paper is organized as follows: In Section~\ref{sec:methods}, we describe the methodology of the work. In Section~\ref{sec:results}, we present the results. Finally, in Section~\ref{sec:conclusion}, we summarize our findings and discuss their implications for AI's capability to assist scientific research.

\section{Methods}
\label{sec:methods}
This work focuses on AI's capability in assisting with one of the main stages of the scientific workflow of a researcher in physics: literature review. To evaluate it, we have conducted a controlled study of eight projects at the frontier of research in physics, astrophysics, and cosmology, in which every project was analyzed under the same standardized prompt and workflow. In the following, we will first give a brief outline of each of the projects, and then describe the study in detail, as well as the procedures to analyze the AI-generated candidate references and their comparison with the bibliography independently selected by a human.

\subsection{The outline of eight research projects}
\label{sec:projects}

The eight expert-conceived projects are outlined below.

\begin{table*}
\renewcommand{\arraystretch}{1.5}
\begin{tabular}{|l|l|l|}
\hline
\parbox[t]{1cm}{\raggedright \textbf{ID}} &
\parbox[t]{8.5cm}{\raggedright \textbf{Project title}} &
\parbox[t]{5cm}{\raggedright \textbf{Domain}} \\
\hline
\parbox[t]{1cm}{\raggedright AGN} &
\parbox[t]{8.5cm}{\raggedright MaNGA: AGN Duty Cycle} &
\parbox[t]{5cm}{\raggedright AGN / galaxy evolution} \\
\hline
\parbox[t]{1cm}{\raggedright LBG} &
\parbox[t]{8.5cm}{\raggedright The galaxy--dark matter halo connection of Lyman-break galaxies} &
\parbox[t]{5cm}{\raggedright Large-scale structure} \\
\hline
\parbox[t]{1cm}{\raggedright IA} &
\parbox[t]{8.5cm}{\raggedright Intrinsic alignments in varying environments} &
\parbox[t]{5cm}{\raggedright Weak lensing / intrinsic alignments} \\
\hline
\parbox[t]{1cm}{\raggedright AR} &
\parbox[t]{8.5cm}{\raggedright Prediction of debris emergence on laser-ablated sub-wavelength shapes for millimeter-wave anti-reflection} &
\parbox[t]{5cm}{\raggedright Instrumentation / laser fabrication} \\
\hline
\parbox[t]{1cm}{\raggedright RG} &
\parbox[t]{8.5cm}{\raggedright Radio Galaxies with HalfDome} &
\parbox[t]{5cm}{\raggedright CMB foregrounds / radio galaxies} \\
\hline
\parbox[t]{1cm}{\raggedright GW} &
\parbox[t]{8.5cm}{\raggedright Studying the environment of gravitational-wave black hole binaries with weak lensing maps} &
\parbox[t]{5cm}{\raggedright Gravitational waves / multi-messenger} \\
\hline
\parbox[t]{1cm}{\raggedright PTA} &
\parbox[t]{8.5cm}{\raggedright Forecasting the pulsar timing array sensitivity needed to measure deviations from general relativity} &
\parbox[t]{5cm}{\raggedright Gravitational waves / tests of GR} \\
\hline
\parbox[t]{1cm}{\raggedright SU2} &
\parbox[t]{8.5cm}{\raggedright Massive Yang--Mills theory (SU(2))} &
\parbox[t]{5cm}{\raggedright Theoretical cosmology / inflation} \\
\hline
\end{tabular}
\caption{The eight expert-conceived research projects used in this study. Titles are as written by the human experts. Each project was independently turned into one human-written and three AI-written one-page proposals.}
\label{tab:projects}
\end{table*}

\textbf{AGN  --  MaNGA: AGN Duty Cycle.} Active galactic nuclei (AGN) are expected to spend $\sim\!10^{6}$--$10^{9}$ year (yr) in the active phase, switching on and off on $\sim\!100$ kyr timescales so that their luminosity fluctuates by orders of magnitude. Because light takes $\sim\!30$ kyr to cross a $\sim\!10$ kpc galaxy, the ionization state of the host disk encodes the recent AGN luminosity history. The project proposes a systematic search for fading or otherwise variable AGN in the MaNGA DR17 survey, with the goal of measuring the fraction of fading/brightening AGN -- i.e., the duty-cycle fraction in the transitional phase -- characterizing the UV luminosity evolution on kyr timescales, and comparing the result to cosmological-simulation predictions.

\textbf{LBG  --  The galaxy--dark matter halo connection of Lyman-break galaxies.} Lyman-break galaxies (LBGs) are identified by the characteristic ``drop-out'' of broadband flux blueward of the Lyman limit, a robust selection for galaxies at $z>2$ that upcoming surveys such as `Ōnohi`ula Prime Focus Spectrograph Galaxy Evolution Survey will exploit to measure redshifts for thousands of LBGs at $2.0<z<4.0$. In preparation for modeling the LBG two-point correlation function, the project will use the Uchuu--UniverseMachine mock catalog to study how LBG-selected samples trace large-scale structure via galaxy bias and the halo occupation distribution as a function of redshift, with particular attention to assembly bias and selection effects that could bias inferences of the galaxy bias and growth rate $f\sigma_8$.

\textbf{IA  --  Intrinsic alignments in varying environments.} The intrinsic alignment (IA) of galaxy shapes is a key contaminant of weak-lensing surveys. IA is known to depend on galaxy color and luminosity; this project investigates an additional, suspected dependence on the large-scale environment (matter density field). The goal is to quantify the dependence of IA on the large-scale density field and to estimate the impact on cosmological analyses if this dependence is neglected.

\textbf{AR  --  Prediction of debris emergence on laser-ablated sub-wavelength shapes.} Sub-wavelength structures (SWS) for millimeter-wave anti-reflection coatings can be machined on hard ceramics using ultra-short-pulse laser ablation. Higher laser power increases the ablation rate but also causes redeposition of ablated debris, the physics of which is poorly understood and whose onset conditions are ambiguous. The goal is to determine the conditions under which debris begins to emerge as a function of arbitrary laser parameters.

\textbf{RG  --  Radio Galaxies with HalfDome.} At CMB frequencies near $100$ GHz, high-energy radio galaxies act as bright point-source contaminants whose positions correlate with the underlying large-scale structure and with other wavelengths (CIB, radio continuum, X-ray). Building on the HalfDome simulations \cite{halfdome}, the project will assign realistic radio luminosities to halos in N-body simulations for CMB foreground modeling, correlated with halo mass and, where needed, across wavelengths.

\textbf{GW  --  Environment of gravitational-wave black hole binaries with weak-lensing maps.} More than 300 binary black hole (BBH) mergers have been detected via gravitational waves, yet their formation environments remain poorly constrained. The project will cross-correlate publicly available GW event localizations with weak-lensing maps to reveal what cosmic environments GW sources prefer, and to estimate the bias of the GW-source distribution relative to the underlying matter distribution.

\textbf{PTA  --  Forecasting pulsar timing array sensitivity to deviations from general relativity.} Pulsar timing arrays (PTAs) measure correlations in pulse arrival-time perturbations between pulsar pairs, encoded in the overlap reduction function (ORF), which provides a sensitive probe of general relativity (GR) versus modified gravity. Focusing on the gravitational-wave phase velocity, the project will forecast the measurement precision required to distinguish $20\%$, $10\%$, and $1\%$ deviations from GR at specified confidence levels, and estimate how many years of observation are needed to reach it.

\textbf{SU2  --  Massive Yang--Mills theory.}
In early-universe cosmology, vector fields have been considered both as fields coupled to the inflaton and as drivers of inflation through self-interacting terms. Motivated by the recently established smooth massless limit of massive Yang--Mills theory with a mass added by hand \cite{Hell:2021oea}, this project investigates whether a self-interacting massive vector field can drive inflation, beginning with the Proca theory with self-interactions and then generalizing to the massive Yang--Mills fields.


\subsection{Participant roles}

This study involved two groups of participants:

\textbf{Human experts.} For each of the eight projects, one human expert -- a graduate student or postdoc who works on that topic as part of their routine research -- conceived and outlined the project, prepared its background and goal, carried out the human literature search, and evaluated the relevance of the AI-generated candidate references\footnote{Graduate students also received guidance from their supervisors in developing their projects.}.

\textbf{AI prompters.} The AI-generated candidate references were produced by a separate group of AI prompters -- undergraduate and graduate students with backgrounds in physics, astrophysics, and cosmology, but without prior experience in the specific projects analyzed here. They issued the standardized prompts, gathered and organized the resulting references, and performed initial checks for hallucinations.

\subsection{Literature workflow}
\label{sec:lw}

The literature review workflow for each of the projects consisted of two main parts: literature search performed by the human, and one generated by AI. Each reference was categorized into one of four classes:
\begin{enumerate}[noitemsep, nolistsep]
    \item \textbf{Review papers}: reviews on the topic from the past 15 years.
    \item \textbf{Highly cited papers}: e.g., the first theoretical proposal or the discovery paper.
    \item \textbf{Recent papers}: results on the topic from the past 10 years.
    \item \textbf{Other papers}: e.g., collaboration papers that set the stage of the topic, such as Planck, WMAP, DES, HSC, or KIDS.
\end{enumerate}

Each human expert conducted a conventional literature search for their project. From all references found, they selected up to 50 that were most important for their study and categorized each into one of the four classes defined above.
In parallel with the human search, and independently of it, the AI-generated candidate references were produced as follows. First, each of the human experts prepared two paragraphs for their corresponding project, the first being the background of the project, and the second being the goal -- \textit{i.e.}, the question that this project is trying to address. Each paragraph had at most three sentences, with all of the material summarized in Appendix~\ref{sec:paragraphs}.
This material was subsequently used by the AI prompters, who prepared it to prompt the LLMs according to the standardized prompt given in Appendix~\ref{app:lit_rev_prompt}. Together with the human experts, they then recorded all of the literature.

The LLMs used primarily in this study to generate candidate references were mid-2025 models, ChatGPT-4o, ChatGPT Deep Research\footnote{ChatGPT Deep Research refers to the tool-augmented research mode available in mid-2025, rather than a separately versioned model.}, and Gemini, with each of them having the online search capability. The literature search using these three LLMs was performed for each of the projects independently. Candidate references were specified according to the following information: Title, Author, Year, Journal, DOI, and Link. Similarly to the search conducted by humans, each individual LLM was capped at 50 papers. They were instructed to prioritize ADS, INSPIRE, and arXiv links, with other links being acceptable if the former were unavailable.

\subsection{Reference comparison}
All of the references are classified into categories listed in the previous subsection, and divided among the human, ChatGPT-4o, ChatGPT Deep Research, and Gemini contributions. 
Once all the literature was obtained, the human experts went through the references generated by AI models and checked the relevance of each reference for the project, marking those that are irrelevant. This choice was necessarily subjective because each person has a different opinion on how relevant a given work is.

We characterize the combined set of references chosen by humans and by individual LLM models as:
\begin{itemize}[noitemsep, nolistsep]
\item Total number of papers found by the humans.
\item Total number of papers generated by the AI.
\item Number of papers found by the humans but missed by the AI.
\item Number of relevant papers found by the AI but missed by the humans.
\item Number of papers found by the AI but deemed irrelevant by the humans.
\end{itemize}

One should note that there were several cases in which an LLM generated a reference with the same title, but placed in two different categories, such as review and highly cited. We have counted such cases as two separate references, generated by a single LLM. This is natural because while some references may be matching in title, they do not necessarily match in the other information, such as their year of publication, journal or DOI. This is also important for the references found by humans and AI. In particular, we define such references as those that match in both the title and the category in which they are placed. For example, if we encounter two references that match in a title, but a human placed one as recent, while the LLM placed it as other, then these are counted as two different references.

The overall results are presented in subsection~\ref{sec:litR}.
\subsection{Hallucination evaluation}
\label{sec:Halev}
In addition to comparing references chosen by humans and LLMs, we performed a comprehensive study of the bibliographic output of AI, assessing to what extent the generated references are hallucinated. We use \textit{hallucination} as an umbrella term for two failure modes of an AI-generated reference: \textit{(i) fabrication}, where the reference points to a nonexistent paper; and \textit{(ii) metadata mismatch}, where the paper exists but one or more of its generated fields (title, first author, year, journal, DOI, or link) disagree with the true values.
The generated references are characterized by the following values: title, authors, year, journal, DOI, and a corresponding link. However, not all references are perfectly generated. For example, a reference's DOI and link may lead to different papers.
To classify this, we considered all of the references that are generated by AI, and not found by a human. We first searched for a reference online based on its generated DOI, and then compared the title, name of the first author, year and the journal of the real reference returned by the Crossref with the AI-generated values. From this, we computed the number of perfect references, for which all of the values match, as well as the number of mismatches, for which some of the values disagree, either partially or completely.

Some of the AI-generated references do not have a working or existing DOI. For these, we retrieved the real reference's DOI from the provided link and compared the values obtained from Crossref with the AI-generated ones.
For the references where the link was not working, we performed an online search based on the title, with which we verified if the resulting reference exists, or is a complete fabrication.

\begin{figure}[ht]
\centering
\resizebox{0.82\columnwidth}{!}{%
\begin{tikzpicture}[
  font=\footnotesize, >=Latex,
  start/.style={draw, rounded corners, fill=gray!8, align=center, text width=4.6cm, inner sep=4pt},
  dec/.style={draw, diamond, aspect=2.4, fill=blue!6, align=center, inner sep=1pt, text width=2.4cm},
  res/.style={draw, rounded corners, fill=green!8, align=center, inner sep=3pt, text width=2.7cm},
  fab/.style={draw, rounded corners, fill=red!8, align=center, inner sep=3pt, text width=2.4cm},
]
\node[start] (s) {641 AI-generated references not found by a human};
\node[dec, below=7mm of s] (doi) {DOI resolves?};
\node[dec, below=15mm of doi] (link) {Link resolves?};
\node[dec, below=15mm of link] (title) {Paper found by title?};
\node[fab, below=9mm of title] (fab) {Fabrication};
\node[res, right=1.9cm of doi] (o1) {Perfect or metadata mismatch};
\node[res, right=1.9cm of link] (o2) {Perfect or metadata mismatch};
\node[res, right=1.9cm of title] (o3) {Metadata mismatch (broken DOI/link)};
\draw[->] (s) -- (doi);
\draw[->] (doi) -- node[left]{no} (link);
\draw[->] (link) -- node[left]{no} (title);
\draw[->] (title) -- node[left]{no} (fab);
\draw[->] (doi) -- node[above]{yes} (o1);
\draw[->] (link) -- node[above]{yes} (o2);
\draw[->] (title) -- node[above]{yes} (o3);
\end{tikzpicture}}
\caption{The hallucination-evaluation procedure. Each AI-generated reference not found by a human is checked first against its DOI, then (as a fallback) against its link, and finally against a title search. A reference whose DOI or link resolves is scored as perfect or as a metadata mismatch by comparing fields with Crossref; one confirmed to exist only by title is a metadata mismatch (its DOI and link are broken); one not found at all is a fabrication.}
\label{fig:halflow}
\end{figure}
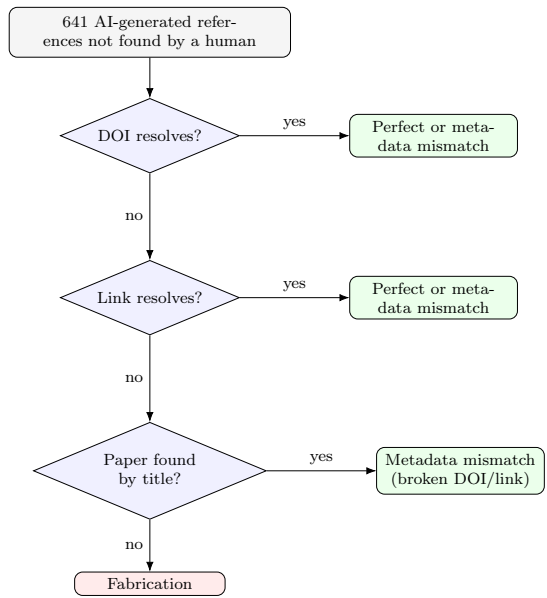

These three checks are applied in sequence, each as a fallback for the previous one, and every reference ends up in exactly one of three outcomes: perfect, metadata mismatch, or fabrication. This procedure is summarized in Fig.~\ref{fig:halflow}.

The overall results are presented in subsection~\ref{sec:Hal}.

\section{Results}
\label{sec:results}
In the following, we will report the analysis of the literature review for the eight research projects in physics, astrophysics and cosmology. First, we will compare the literature selection by human experts through a standard AI-independent search with the AI-generated one. Then, we will present the quality and completeness of the references generated by AI. Both of these sets of results will be based on the mid-2025 LLM models: ChatGPT-4o, ChatGPT Deep Research, and Gemini. Finally, we will present the results for ChatGPT Pro 5.5, which is one of the latest frontier AI models at the time of writing this paper.

\subsection{Literature review: Human vs. AI }
\label{sec:litR}

To generate candidate references,
each LLM was prompted by default in the same way, following the instructions given in Appendix~\ref{app:lit_rev_prompt}. It classified papers as outlined in subsection~\ref{sec:lw} into review papers, highly cited papers, recent papers, or other papers such as collaboration papers that set the stage of the topic.
Interestingly, we have noticed that the same paper may appear twice for a single LLM, classified into two different categories. For example, ChatGPT-4o generated a reference that was presented both as a review paper, and as a highly cited paper. In this case, to analyze the data we have counted this as two separate references due to the different category. Curiously, for the previous ChatGPT-4o example, ChatGPT Deep Research classified the same paper into a third category, recent papers.
The total number of papers found by the human experts was 194. The total number of AI-generated references was 701, with ChatGPT-4o generating 219, ChatGPT Deep Research generating 157, and Gemini generating 325 papers. Each of the three models was run on all eight projects, with up to 50 references per project.

\begin{figure*}[bt]
\centering
\begin{tikzpicture}
\begin{axis}[
    ybar stacked, bar width=17pt,
    width=17.6cm, height=8.4cm,
    ymin=0, ymax=230, ytick={0,50,100,150,200},
    xmin=0.6, xmax=6.2,
    xtick={1.0,1.3,1.6, 2.4,2.7,3.0, 3.8,4.1,4.4, 5.2,5.5,5.8},
    xticklabels={4o,DR,Gm, 4o,DR,Gm, 4o,DR,Gm, 4o,DR,Gm},
    xtick style={draw=none},
    extra x ticks={1.3,2.7,4.1,5.5},
    extra x tick labels={Review, Highly cited, Recent, Other},
    extra x tick style={major tick length=0pt, tick style={draw=none},
                        tick label style={yshift=-15pt, font=\small}},
    ylabel={\# papers},
    tick label style={font=\footnotesize}, label style={font=\small},
    axis line style={gray!50},
    ymajorgrids=true, grid style={dashed, gray!25, line width=0.4pt},
    legend style={at={(0.5,-0.16)}, anchor=north, legend columns=4, font=\small, draw=gray!40},
    legend cell align=left,
]
\addplot+[ybar, fill=llmgreen, draw=black!45] coordinates {(1.0,3)(1.3,0)(1.6,1)(2.4,7)(2.7,12)(3.0,4)(3.8,7)(4.1,6)(4.4,6)(5.2,0)(5.5,0)(5.8,0)};
\addplot+[ybar, fill=llmpink, draw=black!45]  coordinates {(1.0,9)(1.3,12)(1.6,11)(2.4,44)(2.7,39)(3.0,47)(3.8,95)(4.1,96)(4.4,96)(5.2,29)(5.5,29)(5.8,29)};
\addplot+[ybar, fill=llmblue, draw=black!45]  coordinates {(1.0,35)(1.3,30)(1.6,26)(2.4,32)(2.7,27)(3.0,42)(3.8,31)(4.1,38)(4.4,56)(5.2,20)(5.5,24)(5.8,33)};
\addplot+[ybar, fill=gray!55, draw=black!45]  coordinates {(1.0,18)(1.3,6)(1.6,24)(2.4,12)(2.7,7)(3.0,19)(3.8,22)(4.1,6)(4.4,56)(5.2,32)(5.5,1)(5.8,58)};
\legend{{Human \& AI},{Human, not AI},{AI, not human},{AI, irrelevant}}
\end{axis}
\end{tikzpicture}
\caption{\label{fig:litrevfig}
Composition of the combined human$+$AI reference set for each model, split by literature type. Within each literature type (Review, Highly cited, Recent, Other) the three bars are ChatGPT-4o (4o), ChatGPT Deep Research (DR), and Gemini (Gm); bar height is the number of papers, stacked into: found by both the human expert and the AI (Human \& AI, green); found by the human but missed by the AI (Human, not AI, pink); found by the AI and judged relevant but missed by the human (AI, not human, blue); and found by the AI but judged irrelevant (AI, irrelevant, grey). Exact counts and percentages are in Table~\ref{tab:LLMoverall}.}
\end{figure*}
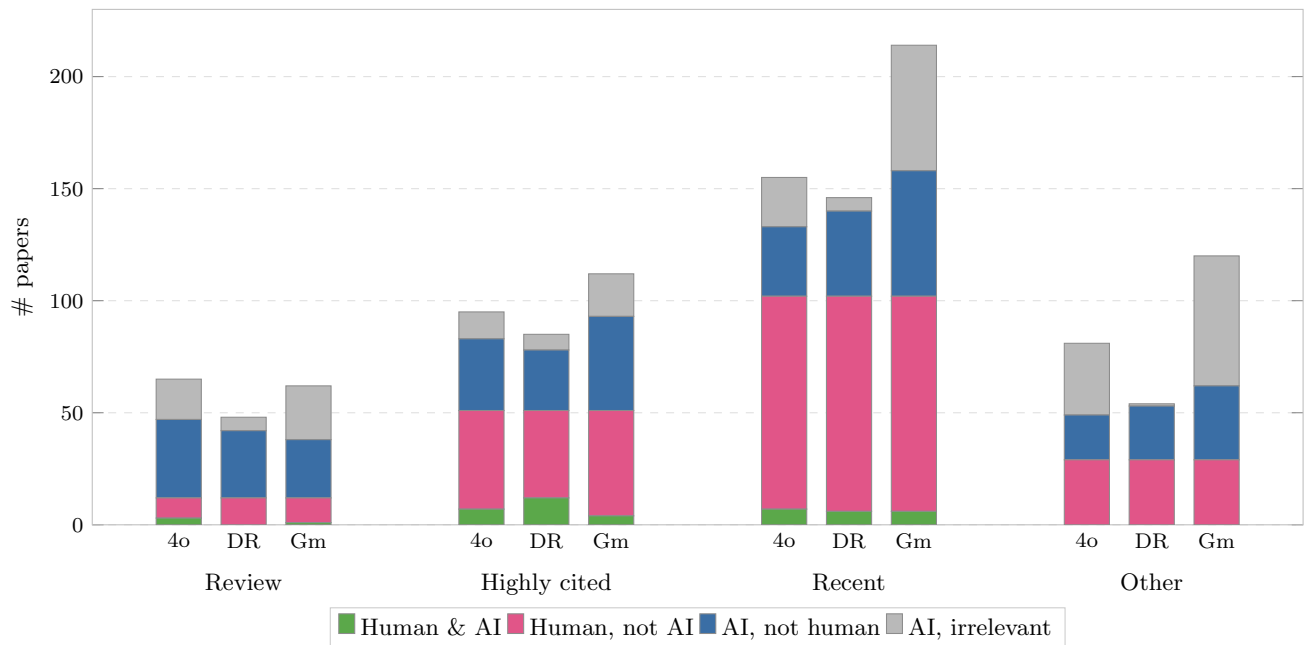

For each model, and separately for each of the four literature types, we compared its references with the human selection and classified every reference into one of four groups: found by both the human expert and the AI; found by the human but missed by the AI; found by the AI and judged relevant but missed by the human; and found by the AI but judged irrelevant. The resulting composition is shown in Figure~\ref{fig:litrevfig}, with the underlying counts and percentages given in Table~\ref{tab:LLMoverall}.

\begin{table*}[ht]
\centering
\begin{tabular}{llccccr}
\toprule
\textbf{LLM} & \textbf{Category} & \textbf{Rev.} & \textbf{H.\,cited} & \textbf{Recent} & \textbf{Other} & \textbf{Total (\%)} \\
\midrule
\multirow{4}{*}{ChatGPT-4o}
 & Human \& AI     & 3  & 7  & 7  & 0  & 17 (4.3\%) \\
 & Human, not AI   & 9  & 44 & 95 & 29 & 177 (44.7\%) \\
 & AI, not human   & 35 & 32 & 31 & 20 & 118 (29.8\%) \\
 & AI, irrelevant  & 18 & 12 & 22 & 32 & 84 (21.2\%) \\
\midrule
\multirow{4}{*}{ChatGPT DR}
 & Human \& AI     & 0  & 12 & 6  & 0  & 18 (5.4\%) \\
 & Human, not AI   & 12 & 39 & 96 & 29 & 176 (52.9\%) \\
 & AI, not human   & 30 & 27 & 38 & 24 & 119 (35.7\%) \\
 & AI, irrelevant  & 6  & 7  & 6  & 1  & 20 (6.0\%) \\
\midrule
\multirow{4}{*}{Gemini}
 & Human \& AI     & 1  & 4  & 6  & 0  & 11 (2.2\%) \\
 & Human, not AI   & 11 & 47 & 96 & 29 & 183 (36.0\%) \\
 & AI, not human   & 26 & 42 & 56 & 33 & 157 (30.9\%) \\
 & AI, irrelevant  & 24 & 19 & 56 & 58 & 157 (30.9\%) \\
\bottomrule
\end{tabular}
\caption{Breakdown of the combined human$+$AI reference set (the data plotted in Fig.~\ref{fig:litrevfig}) by literature type and category, for each model. Columns give the counts per literature type (Review, Highly cited, Recent, Other) and the total, with the percentage of that model's combined human$+$AI set in parentheses (union totals of 396, 333, and 508 references for ChatGPT-4o, ChatGPT Deep Research, and Gemini). ``Human \& AI'' = found by both; ``Human, not AI'' = found by the human but missed by the AI; ``AI, not human'' = found by the AI and judged relevant but missed by the human; ``AI, irrelevant'' = found by the AI but judged irrelevant.}
\label{tab:LLMoverall}
\end{table*}
We can notice that the highest percentage for all three LLMs is for papers found by the human experts and not by the AIs. However, the set of papers generated by the LLMs and not found by humans is also non-negligible. For ChatGPT-4o and ChatGPT Deep Research, this fraction is smaller than the papers found by humans and not by the AIs ($30\%$ vs $45\%$, and $36\%$ vs $53\%$, respectively), while for Gemini the two fractions differ by only $5\%$ ($31\%$ vs $36\%$). This suggests that AIs can positively contribute to the literature assessment as part of the scientific workflow of the researcher, and generate papers that would otherwise have been omitted.
Interestingly, while Gemini has generated a substantially higher number of papers, it also has a high rate of irrelevance ($31\%$) when compared to the other two models. ChatGPT-4o still has a relatively high one ($21\%$), but ChatGPT Deep Research has a low rate of only $6\%$.
Following the titles across different projects, AI appears to mainly focus on the key-word search characterizing the problem, which may lead it to select works that are irrelevant to researchers, but also provide a more comprehensive list of topics that are closely related to the project. Overall, references suggested by humans appear to move past these key-words, sometimes focusing also on the foundational work or related but separate fields.
\subsection{Hallucinations}
\label{sec:Hal}

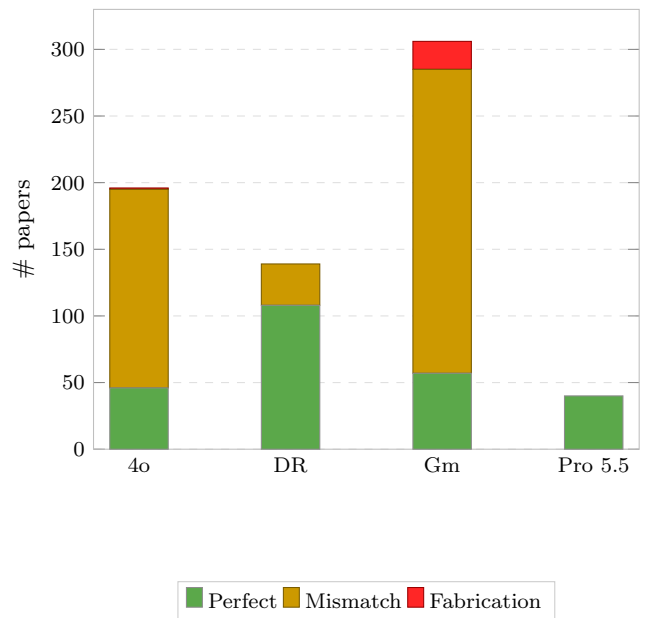
\begin{figure}[ht]
\centering
\begin{tikzpicture}
\begin{axis}[
    ybar stacked, bar width=22pt,
    width=8.8cm, height=7.4cm,
    ymin=0, ymax=330, ytick={0,50,100,150,200,250,300},
    symbolic x coords={4o, DR, Gm, Pro 5.5},
    xtick=data, x tick label style={font=\footnotesize},
    xtick style={draw=none},
    ylabel={\# papers},
    tick label style={font=\footnotesize}, label style={font=\small},
    axis line style={gray!50},
    ymajorgrids=true, grid style={dashed, gray!25, line width=0.4pt},
    legend style={at={(0.5,-0.30)}, anchor=north, legend columns=3, font=\footnotesize, draw=gray!40},
    legend cell align=left,
]
\addplot+[ybar, fill=llmgreen, draw=black!45]      coordinates {(4o,46) (DR,108) (Gm,57) (Pro 5.5,40)};
\addplot+[ybar, fill=darkyellow, draw=darkyellow!60!black]  coordinates {(4o,149) (DR,31) (Gm,228) (Pro 5.5,0)};
\addplot+[ybar, fill=red!85, draw=red!60!black]     coordinates {(4o,1) (DR,0) (Gm,21) (Pro 5.5,0)};
\legend{{Perfect},{Mismatch},{Fabrication}}
\end{axis}
\label{fig:halbars}
\end{tikzpicture}
\caption{Hallucination breakdown per model: number of AI-generated references (not found by a human) that are perfect (green), metadata mismatches (dark yellow), or fabrications (red). ChatGPT-4o, ChatGPT Deep Research, and Gemini are over all eight projects; ChatGPT Pro~5.5 is for the single SU(2) project only (Sec.~\ref{sec:Pro5.5}) and produced only perfect references.}
\end{figure}

The total number of AI-generated references considered in this analysis -- those not found by a human -- is 641. One should note that this number is smaller than in the previous subsection because here we do not distinguish between the different categories into which the same paper may have been placed. Following the verification procedure of subsection~\ref{sec:Halev} (summarized in Fig.~\ref{fig:halflow}), each reference is classified as a \textit{perfect} reference (all fields match the true paper), a \textit{metadata mismatch} (a real paper with at least one incorrect field), or a \textit{fabrication} (a reference to a nonexistent paper). The overall results are summarized in Table~\ref{tableGenSpec}.
\begin{table}[ht]
\centering
\begin{tabular}{lrrrrr}
\toprule
 & \textbf{Total} & \textbf{4o} & \textbf{DR} & \textbf{Gemini} & \textbf{Pro 5.5$^{a}$} \\
\midrule
AI-generated references & 641 & 196 & 139 & 306 & 40 \\
Perfect                 & 211 & 46  & 108 & 57  & 40 \\
Fabrication             & 22  & 1   & 0   & 21  & 0  \\
Metadata mismatch       & 408 & 149 & 31  & 228 & 0  \\
\bottomrule
\end{tabular}
\caption{Hallucination evaluation of the AI-generated references not found by humans, overall (Total) and per model (4o~$=$~ChatGPT-4o, DR~$=$~ChatGPT Deep Research). ``Perfect'' $=$ all fields match the true paper; ``Fabrication'' $=$ a nonexistent paper; ``Metadata mismatch'' $=$ a real paper with at least one incorrect field. $^{a}$ChatGPT Pro~5.5 was run for a single project (SU(2)) only (Sec.~\ref{sec:Pro5.5}); its column is not comparable in scale and is excluded from the Total.}
\label{tableGenSpec}
\end{table}
Of the 641 references, 211 (33\%) are perfect, 22 (3\%) are fabrications, and 408 (64\%) are metadata mismatches. The small fabrication rate but large mismatch rate indicates that AI-generated references from the mid-2025 models should be taken with caution, following the trend of the past years. Table~\ref{tablemismatch} breaks down which fields disagree for the 399 mismatched references whose true paper could be resolved through a working DOI or link; the remaining 9 mismatches are real papers with a broken DOI and link, confirmed only through a title-based search, and so could not be checked field by field. The mismatches occur mostly in the first author (306) and title (181), with year (105) and journal (124) less frequent, and most often two of the four fields are wrong (206).
\begin{table}[ht]
\centering
\begin{tabular}{lc}
\toprule
 \textbf{Mismatch type} & \textbf{\# papers} \\
\midrule
Title & 181\\
First author & 306 \\
Year & 105 \\
Journal & 124\\
\midrule
1/4 mismatch & 102 \\
2/4 mismatch & 206 \\
3/4 mismatch & 37 \\
4/4 mismatch & 54 \\
\bottomrule
\end{tabular}
\caption{Types of field mismatch for the 399 mismatched references whose true paper could be resolved through a working DOI or link. Top block: number of mismatches per field. Bottom block: how many of the four fields (title, first author, year, journal) disagree.}
\label{tablemismatch}
\end{table}
The per-model breakdown is included in Table~\ref{tableGenSpec} and visualized in Fig.~\ref{fig:halbars}.
Overall, ChatGPT Deep Research is clearly the most reliable: it produced no fabrications and the highest fraction of perfect references (78\%). Gemini generated both the most fabrications (21) and the most mismatches, while ChatGPT-4o produced only a single fabrication but a large fraction of mismatches (76\%). This ordering mirrors the aggregate picture: the model with stronger tool use and verification (Deep Research) hallucinates markedly less than the plain chat models.
\subsection{How far did we get? Into the ChatGPT Pro 5.5}
\label{sec:Pro5.5}

Our analysis is mostly focused on the mid-2025 models, when our controlled study took place. However, in just a couple of months, LLMs can have several upgrades. To evaluate how this reflected on the completeness of the generated references, we have made a comparison of the ChatGPT Pro 5.5 (dated July 2nd, 2026) and ChatGPT-4o, with focus on the SU(2) project.

To generate the references, we used the same prompts and procedure as outlined in Section~\ref{sec:methods}, with a cap of 50 papers. The total number of papers generated by ChatGPT Pro 5.5 was 44, with an overlap of 4 papers with the references selected by the human expert. In contrast, the ChatGPT-4o version generated 37 substantially different papers, with 4 overlapping with literature selected by the human expert. One can say that the improvement for the newer version is immediate: all references generated by the Pro version are perfect, with all values (title, author, year, journal) matching to perfect accuracy. The ChatGPT Pro 5.5 result is shown alongside the mid-2025 models in Fig.~\ref{fig:halbars}. Strictly speaking, one should note that ChatGPT-4o found 42 different papers, 5 of which refer to the same work, but differ in the link, or DOI, and hence should be counted as different references. We have omitted these near-duplicate references and considered only substantially different ones, while we have counted them in the previous subsections. In contrast, all references generated by ChatGPT Pro 5.5 are substantially different.

It is worth noting that the overlap between the references selected by the human expert and generated by ChatGPT-4o and ChatGPT Pro 5.5 models is given by six different papers, with four overlaps in each model\footnote{The two LLMs share two of these overlaps, while each has two more that the other does not.}. In addition, among these, two overlap between ChatGPT-4o and ChatGPT Pro 5.5. The overall overlap between the ChatGPT-4o and ChatGPT Pro 5.5 is 7 papers. While this number is relatively small, it should be noted that the titles of references in which they do not overlap are similar.
For example, the human expert's selection leaned toward foundational field-theory papers on Yang--Mills gauge fields (with and without a mass term) and their non-linear interactions, along with cosmological applications such as inflation driven by such fields. However, while the possible cap was at 50 papers, only 14 were selected.
ChatGPT-4o primarily focused on the inflationary models connected with the Abelian and non-Abelian vector field modes, including also the stability analysis of models with massive vector fields such as the generalized Proca theory. This trend has also continued in the literature generated by ChatGPT Pro 5.5. In contrast to the human expert, neither LLM selected references with a substantial departure from the cosmological models. This is interesting as it reflects the subjectivity of the literature search. The human expert might have searched more broadly, into the quantum field theory aspects beyond the core of the problem due to the original motivation of the project, although this remains a speculation based on the observed data. In addition, it also reflects the results on the literature search for the mid-2025 LLMs when compared to the human literature selections. It appears AIs are more key-word driven while humans do not necessarily follow that rule.
\section{Conclusion}
\label{sec:conclusion}
We performed a broad, controlled 
 study to test AI's capability to assist with one of the key components of a scientific workflow in the literature review. This was done across eight different projects in physics, astrophysics and cosmology, conceived by experts in the field, with the literature search carried out by the human experts as part of their research and, in parallel, generated by the AI prompters.
For each of the projects, we compared the literature selected by humans to the references generated by LLMs. We studied the relevance of the AI-generated candidate references with respect to each of the projects, as well as its completeness and reliability, focusing on whether the citation values were perfectly accurate or fabricated.
Most of these results concern the mid-2025 LLMs, ChatGPT-4o, ChatGPT Deep Research, and Gemini. To assess how much AI has progressed, we performed a literature search also using ChatGPT Pro 5.5 for one of the projects.
In the following, we summarize our key findings:
\begin{enumerate}
 \item Both AI and human experts generated a large set of relevant references that were essentially not overlapping, indicating that a collaboration between humans and AI may yield a more complete reference overview.
 \item For the mid-2025 models, 3\% of the references are fabrications (nonexistent papers), while 64\% are real papers with at least one incorrect field (a metadata mismatch); both are forms of hallucination. This is one of the main weaknesses of AI, which has persisted in recent years in scientific literature searches, and suggests that one should carefully check the generated references. However, it should be noted that among these, ChatGPT Deep Research is much more reliable than the other two, as it generated no fabricated references, and a smaller number of metadata mismatches corresponding to 22\%. Notably, for the SU(2) project, ChatGPT Pro 5.5 generated all of the references with fully accurate values, indicating an improvement over the previous ChatGPT models. This likely reflects stronger tool use and verification in these models rather than the language model alone, so the reduction in these inaccuracies should be read as a property of the full model-plus-tools system.


 \item AIs appear to prefer more specialized search guided by the project keywords, focused narrowly on the project topic, while human experts appear to search more broadly, taking into account also related fields or more fundamental papers.
\end{enumerate}
Our findings also come with two important caveats. First, we have limited our investigation to only eight projects, while a more comprehensive study could include a larger set. Second, the comparison of the ChatGPT-4o and ChatGPT Pro 5.5 was performed only for a single project.
One should note that it takes a substantial amount of coordination to keep the projects on an equal timeline. This is reflected in our study, which already covers the research frontier across several observational, instrumental, numerical and theoretical projects in physics, astrophysics and cosmology. Therefore, our results should be taken as a snapshot in time, collecting eight groups of people, setting them against AI, and performing a real-life comparison while asking the question -- \textit{Can LLMs help physicists in their scientific research?}

Moreover, while our result on the improvement of AI performance in metadata completeness is based on a relatively small set, it is without a doubt an intriguing result: one of the major weaknesses of AI is hallucinations. If this reduction in fabricated references holds up under a systematic benchmark, it would mark meaningful progress -- though, as noted above, it likely reflects improved tool use and greater test-time compute rather than a change in the underlying language model alone. While our results show an indication that this takes place, a comprehensive benchmark on the physics literature and even beyond would be ideal.
Our study focused on AI's capability in performing literature review -- one of the main steps in a scientific workflow of a physicist. In a companion paper, we will present the results also for the project planning, another major step in research. In future studies, we also hope to carry out a dedicated study of AI's capability in executing projects and writing papers in areas of physics, astrophysics and cosmology. Notably, a recent study indicates that having an LLM carry out the research is promising, at least within the context of algorithmic theoretical physics with sufficient context \cite{Hell:2026dqx}. Nevertheless, already in the first stage of research, we can see that AI shows promise to be a good collaborator in the literature search.



\begin{acknowledgments}
We thank Jingjing Shi, Qiuyue Liang, and Kenta Hotokezaka for help shaping some of the scientific projects, and Daniela Breitman for useful suggestions.
JL acknowledges support from the Kavli Foundation and Google. KV acknowledges support from the Kavli Foundation. AH was supported in part by JSPS KAKENHI Grant No.~JP26K17133, the CD3 Google Seed grant, and by the World Premier International Research Center Initiative (WPI), MEXT, Japan. KK acknowledges support from the Forefront Physics and Mathematics Program to Drive Transformation (FoPM), a World-leading Innovative Graduate Study (WINGS) Program, the University of Tokyo. JG and ZL acknowledge support from the International Laboratory for Astrophysics, Neutrino and Cosmology Experiments (ILANCE). AEB is supported by the Simons Foundation. MRG acknowledges financial support from the Formaci\'on del Profesorado Universitario program of the Spanish Ministerio de Ciencia, Innovaci\'on y Universidades; from the CMB-Inflate project funded by the European Union's Horizon~2020 Research and Innovation Staff Exchange under the Marie Sk{\l}odowska-Curie grant agreement No.~101007633; and from MICIU/AEI/10.13039/501100011033 under projects PID2022-139223OB-C21 and PID2022-140670NA-I00 (also funded by FEDER, UE).
LT is supported by JSPS under KAKENHI 24K22878 and 26K17136 and by the Royal Society under ICA\textbackslash R2\textbackslash 252140.

AI usage: Candidate references were generated using three assistants -- ChatGPT-4o (OpenAI), ChatGPT Deep Research, and Gemini (Gemini-2.5-flash, the default version) -- during mid-2025, with an additional single-project test using ChatGPT Pro 5.5 (dated July 2026). Beyond their role as study instruments, LLMs were also used to assist with language editing, figure preparation, and formatting of this manuscript; all scientific content, analyses, and conclusions are the authors' own.
\end{acknowledgments}

\section*{Author affiliations}
{\footnotesize \raggedright \noindent
$^{1}$Center for Data-Driven Discovery, Kavli IPMU (WPI), UTIAS, The University of Tokyo, Kashiwa, Chiba 277-8583, Japan \\
$^{2}$Kavli IPMU (WPI), UTIAS, The University of Tokyo, 5-1-5 Kashiwanoha, Kashiwa, Chiba 277-8583, Japan \\
$^{3}$Department of Astrophysical Sciences, Princeton University, Peyton Hall, Princeton, NJ 08544, USA \\
$^{4}$Department of Physics, The University of Tokyo, 7-3-1 Hongo, Bunkyo-ku, Tokyo 113-0033, Japan \\
$^{5}$Center for Computational Astrophysics, Flatiron Institute, 162 5th Avenue, New York, NY 10010, USA \\
$^{6}$Department of Physics, University of Oxford, Denys Wilkinson Building, Keble Road, Oxford OX1 3RH, United Kingdom \\
$^{7}$Research Center for the Early Universe, The University of Tokyo, Bunkyo-ku, Tokyo 113-0033, Japan \\
$^{8}$\'Ecole polytechnique, Institut polytechnique de Paris, Palaiseau, France \\
$^{9}$Advanced Energy, Graduate School of Frontier Sciences, The University of Tokyo, Kashiwa, Chiba 277-8561, Japan \\
$^{10}$Department of Astronomy, University of Texas at Austin, Austin, Texas, USA \\
$^{11}$Department of Physics, Graduate School of Science, The University of Tokyo, Bunkyo-ku, Tokyo 113-0033, Japan \\
$^{12}$Instituto de F\'isica de Cantabria (IFCA, CSIC--UC), Avenida los Castros s/n, 39005 Santander, Spain \\
$^{13}$Departamento de F\'isica Moderna, Universidad de Cantabria, Avenida los Castros s/n, E-39005 Santander, Spain \\
$^{14}$Institute for Cosmic Ray Research, The University of Tokyo, 5-1-5 Kashiwa-no-Ha, Kashiwa, Chiba 277-8582, Japan \\
$^{15}$Department of Astronomy, University of Science and Technology of China, Hefei, Anhui 230026, People's Republic of China \\
$^{16}$School of Astronomy and Space Sciences, University of Science and Technology of China, Hefei, Anhui 230026, People's Republic of China \\
}
\bibliography{refs}{}

@article{Scherbakov_2025,
   title={The emergence of large language models as tools in literature reviews: a large language model-assisted systematic review},
   volume={32},
   ISSN={1527-974X},
   url={http://dx.doi.org/10.1093/jamia/ocaf063},
   DOI={10.1093/jamia/ocaf063},
   number={6},
   journal={Journal of the American Medical Informatics Association},
   publisher={Oxford University Press (OUP)},
   author={Scherbakov, Dmitry and Hubig, Nina and Jansari, Vinita and Bakumenko, Alexander and Lenert, Leslie A},
   year={2025},
   month=May, pages={1071–1086} }

@ARTICLE{11132297,
  author={Woesle, Christian and Fischer-Brandies, Leopold and Buettner, Ricardo},
  journal={IEEE Access}, 
  title={A Systematic Literature Review of Hallucinations in Large Language Models}, 
  year={2025},
  volume={13},
  number={},
  pages={148231-148253},
  doi={10.1109/ACCESS.2025.3601206}}

@article{Walters_2023,
   title={Fabrication and errors in the bibliographic citations generated by ChatGPT.},
   volume={13},
   url={https://doi.org/10.1038/s41598-023-41032-5},
   DOI={https://doi.org/10.1038/s41598-023-41032-5},
   journal={Sci Rep },
   author={Walters, William H. and Wilder, Esther Isabelle },
   year={2023},
   pages={14045} }

@article{Franzoni_2024,
   title={Retracting ChatGPT: completeness and relevance of academic references. },
   volume={226},
   url={https://doi.org/10.1007/s44217-024-00333-1},
   DOI={https://doi.org/10.1007/s44217-024-00333-1},
   journal={ Discov Educ 3 },
   author={Franzoni Velázquez, Ana Lidia and Huerta, Esperanza and Jensen, Scott},
   year={2024},
   pages={226} }

@article{Pastucha_2025,
   title={Reference Accuracy in Large Language Model Chatbots: A Metric for Inherent Misinformation?.},
   volume={32},
   DOI={10.12659/MSM.950916},
   journal={Medical science monitor : international medical journal of experimental and clinical research},
   author={Pastucha, Małgorzata et al. },
   year={2026},
   pages={e950916} }

@article{Hezaveh2017,
  author  = {Hezaveh, Yashar D. and Perreault Levasseur, Laurence and Marshall, Philip J.},
  title   = {Fast automated analysis of strong gravitational lenses with convolutional neural networks},
  journal = {Nature},
  volume  = {548},
  number  = {7669},
  pages   = {555--557},
  year    = {2017},
  doi     = {10.1038/nature23463}
}

@article{VillaescusaNavarro2021,
  author  = {Villaescusa-Navarro, Francisco and Angl{\'e}s-Alc{\'a}zar, Daniel and Genel, Shy and Spergel, David N. and Somerville, Rachel S. and Dave, Romeel and Pillepich, Annalisa and Hernquist, Lars and Nelson, Dylan and Torrey, Paul and others},
  title   = {The {CAMELS} Project: Cosmology and Astrophysics with Machine-learning Simulations},
  journal = {The Astrophysical Journal},
  volume  = {915},
  number  = {1},
  pages   = {71},
  year    = {2021},
  doi     = {10.3847/1538-4357/abf7ba}
}

@inproceedings{Nguyen2023,
  author    = {Nguyen, Tuan Dung and Ting, Yuan-Sen and Ciuc{\u{a}}, Ioana and O'Neill, Charlie and Sun, Ze-Chang and Jab{\l}o{\'n}ska, Maja and Kruk, Sandor and Perkowski, Ernest and Miller, Jack and Li, Jason and others},
  title     = {{AstroLLaMA}: Towards Specialized Foundation Models in Astronomy},
  booktitle = {Proceedings of the Second Workshop on Information Extraction from Scientific Publications (WIESP), IJCNLP-AACL 2023},
  pages     = {49--55},
  year      = {2023},
  note      = {arXiv:2309.06126}
}

@inproceedings{Si2024,
  author    = {Si, Chenglei and Yang, Diyi and Hashimoto, Tatsunori},
  title     = {Can {LLMs} Generate Novel Research Ideas? A Large-Scale Human Study with 100+ {NLP} Researchers},
  booktitle = {International Conference on Learning Representations (ICLR)},
  year      = {2025},
  note      = {arXiv:2409.04109}
}

@article{Lu2024,
  author  = {Lu, Chris and Lu, Cong and Lange, Robert Tjarko and Foerster, Jakob and Clune, Jeff and Ha, David},
  title   = {The {AI} Scientist: Towards Fully Automated Open-Ended Scientific Discovery},
  journal = {arXiv e-prints},
  year    = {2024},
  note    = {arXiv:2408.06292}
}

@article{Liang2024,
  author  = {Liang, Weixin and Zhang, Yuhui and Cao, Hancheng and Wang, Binglu and Ding, Daisy Yi and Yang, Xinyu and Vodrahalli, Kailas and He, Siyu and Smith, Daniel Scott and Yin, Yian and McFarland, Daniel A. and Zou, James},
  title   = {Can Large Language Models Provide Useful Feedback on Research Papers? A Large-Scale Empirical Analysis},
  journal = {NEJM AI},
  volume  = {1},
  number  = {8},
  year    = {2024},
  doi     = {10.1056/AIoa2400196},
  note    = {arXiv:2310.01783}
}

@inproceedings{Mitchell2023,
  author    = {Mitchell, Eric and Lee, Yoonho and Khazatsky, Alexander and Manning, Christopher D. and Finn, Chelsea},
  title     = {{DetectGPT}: Zero-Shot Machine-Generated Text Detection using Probability Curvature},
  booktitle = {Proceedings of the 40th International Conference on Machine Learning (ICML)},
  series    = {PMLR},
  volume    = {202},
  pages     = {24950--24962},
  year      = {2023},
  note      = {arXiv:2301.11305}
}

@article{Gao2023,
  author  = {Gao, Catherine A. and Howard, Frederick M. and Markov, Nikolay S. and Dyer, Emma C. and Ramesh, Siddhi and Luo, Yuan and Pearson, Alexander T.},
  title   = {Comparing scientific abstracts generated by {ChatGPT} to real abstracts with detectors and blinded human reviewers},
  journal = {npj Digital Medicine},
  volume  = {6},
  number  = {1},
  pages   = {75},
  year    = {2023},
  doi     = {10.1038/s41746-023-00819-6}
}

@article{Shcherbiak2024,
  author  = {Shcherbiak, Anna and Habibnia, Hooman and B{\"o}hm, Robert and Fiedler, Susann},
  title   = {Evaluating science: A comparison of human and {AI} reviewers},
  journal = {Judgment and Decision Making},
  volume  = {19},
  pages   = {e21},
  year    = {2024},
  doi     = {10.1017/jdm.2024.24}
}

@inproceedings{Panickssery2024,
  author    = {Panickssery, Arjun and Bowman, Samuel R. and Feng, Shi},
  title     = {{LLM} Evaluators Recognize and Favor Their Own Generations},
  booktitle = {Advances in Neural Information Processing Systems (NeurIPS)},
  volume    = {37},
  year      = {2024},
  note      = {arXiv:2404.13076}
}

@article{Wataoka2024,
  author  = {Wataoka, Koki and Takahashi, Tsubasa and Ri, Ryokan},
  title   = {Self-Preference Bias in {LLM}-as-a-Judge},
  journal = {arXiv e-prints},
  year    = {2024},
  note    = {arXiv:2410.21819. Presented at the NeurIPS 2024 Safe Generative AI Workshop}
}

@inproceedings{Zhou2024,
  author    = {Zhou, Hongli and Huang, Hui and Long, Yunfei and Xu, Bing and Zhu, Conghui and Cao, Hailong and Yang, Muyun and Zhao, Tiejun},
  title     = {Mitigating the Bias of Large Language Model Evaluation},
  booktitle = {Proceedings of the 23rd Chinese National Conference on Computational Linguistics (CCL)},
  pages     = {1310--1319},
  year      = {2024},
  note      = {arXiv:2409.16788}
}

@article{Sadasivan2023,
  author  = {Sadasivan, Vinu Sankar and Kumar, Aounon and Balasubramanian, Sriram and Wang, Wenxiao and Feizi, Soheil},
  title   = {Can {AI}-Generated Text be Reliably Detected?},
  journal = {arXiv e-prints},
  year    = {2023},
  note    = {arXiv:2303.11156}
}

@article{Ren2025,
  author  = {Ren, Jing and Wang, Weiqi},
  title   = {Assisting Research Proposal Writing with Large Language Models: Evaluation and Refinement},
  journal = {arXiv e-prints},
  year    = {2025},
  note    = {arXiv:2509.09709}
}

@article{Sandstrom2026,
  author  = {Sandstr{\"o}m, Ulf and Thelwall, Mike},
  title   = {Can Large Language Models Evaluate Grant Proposal Quality? Revisiting the Wenner{\aa}s and Wold Peer Review Data},
  journal = {arXiv e-prints},
  year    = {2026},
  note    = {arXiv:2603.14565}
}

@article{Thorne2026,
  author  = {Thorne, William and James, Joseph and Wang, Yang},
  title   = {Evaluating {LLM}-Based Grant Proposal Review via Structured Perturbations},
  journal = {arXiv e-prints},
  year    = {2026},
  note    = {arXiv:2603.08281}
}

@article{Sikimic2025,
  author  = {Sikimi{\'c}, Vlasta},
  title   = {Fair or flawed? Rethinking grant review with generative {AI}},
  journal = {Synthese},
  volume  = {206},
  pages   = {282},
  year    = {2025},
  doi     = {10.1007/s11229-025-05366-z}
}

@article{halfdome,
    author = "Bayer, Adrian E. and Zhong, Yici and Li, Zack and DeRose, Joseph and Feng, Yu and Liu, Jia",
    title = "{The HalfDome multi-survey cosmological simulations: N-body simulations}",
    eprint = "2407.17462",
    archivePrefix = "arXiv",
    primaryClass = "astro-ph.CO",
    doi = "10.1088/1475-7516/2025/05/016",
    journal = "JCAP",
    volume = "05",
    pages = "016",
    year = "2025"
}

@misc{denario,
      title={The Denario project: Deep knowledge AI agents for scientific discovery}, 
      author={Francisco Villaescusa-Navarro and Boris Bolliet and Pablo Villanueva-Domingo and Adrian E. Bayer and Aidan Acquah and Chetana Amancharla and Almog Barzilay-Siegal and Pablo Bermejo and Camille Bilodeau and Pablo Cárdenas Ramírez and Miles Cranmer and Urbano L. França and ChangHoon Hahn and Yan-Fei Jiang and Raul Jimenez and Jun-Young Lee and Antonio Lerario and Osman Mamun and Thomas Meier and Anupam A. Ojha and Pavlos Protopapas and Shimanto Roy and David N. Spergel and Pedro Tarancón-Álvarez and Ujjwal Tiwari and Matteo Viel and Digvijay Wadekar and Chi Wang and Bonny Y. Wang and Licong Xu and Yossi Yovel and Shuwen Yue and Wen-Han Zhou and Qiyao Zhu and Jiajun Zou and Íñigo Zubeldia},
      year={2025},
      eprint={2510.26887},
      archivePrefix={arXiv},
      primaryClass={cs.AI},
      url={https://arxiv.org/abs/2510.26887}, 
}

@misc{Hell:2026dqx,
    author = "Hell, Anamaria and Thiele, Leander",
    title = "{LLMs with in-context learning for Algorithmic Theoretical Physics}",
    eprint = "2605.08212",
    archivePrefix = "arXiv",
    primaryClass = "cs.LG",
    reportNumber = "IPMU26-0020",
    month = "5",
    year = "2026"
}

@article{Hell:2021oea,
    author = "Hell, Anamaria",
    title = "{The strong couplings of massive Yang-Mills theory}",
    eprint = "2111.00017",
    archivePrefix = "arXiv",
    primaryClass = "hep-th",
    doi = "10.1007/JHEP03(2022)167",
    journal = "JHEP",
    volume = "03",
    pages = "167",
    year = "2022"
}

@misc{Miao:2026gmx,
    author = "Miao, Tingjia and others",
    title = "{PRL-Bench: A Comprehensive Benchmark Evaluating LLMs' Capabilities in Frontier Physics Research}",
    eprint = "2604.15411",
    archivePrefix = "arXiv",
    primaryClass = "cs.LG",
    month = "4",
    year = "2026"
}

@article{2026arXiv260121165W,
       author = {{Wang}, Miles and {Lin}, Robi and {Hu}, Kat and {Jiao}, Joy and {Chowdhury}, Neil and {Chang}, Ethan and {Patwardhan}, Tejal},
        title = "{FrontierScience: Evaluating AI's Ability to Perform Expert-Level Scientific Tasks}",
      journal = {arXiv e-prints},
         year = 2026,
        month = jan,
          eid = {arXiv:2601.21165},
        pages = {arXiv:2601.21165},
          doi = {10.48550/arXiv.2601.21165},
archivePrefix = {arXiv},
       eprint = {2601.21165},
 primaryClass = {cs.AI},
       adsurl = {https://ui.adsabs.harvard.edu/abs/2026arXiv260121165W}
}

@misc{Paper2,
    author = "Liu, Jia and Krishnaraj, Veena and Vovk,  Kateryna and and Aizawa, Kosuke and Bayer, Adrian E. and Blot, Linda and Cowell, Jessica and Garg, Suyog and Gr\'ee, Jonathan and Hell, Anamaria and Horowitz, Ben and Ichikawa, Masaya and Iemoto, Kanyuni and Kondo, Keigo and Lorsin, Zacharie and McCarthy, Kevin and Robinson, Jamie and Ruiz-Granda, Miguel and Thiele, Leander and Vovk, Ievgen and Zhou, Mingshen",
    title = "{AI's Capability in Assisting Scientific Research in Physics, Astrophysics, and Cosmology II: Project Planning and Proposal Evaluation}",
    eprint = "2607.xxxx",
    archivePrefix = "arXiv",
}
\bibliographystyle{apsrev4-2}
\appendix
\section{Background and goals of the eight research projects}
\label{sec:paragraphs}
The following project titles, backgrounds, and goals were written by the human experts without any AI assistance, and were the common input provided to the human planners and the AI prompters for each project (Sec.~\ref{sec:projects}).
\subsection{AGN  --  MaNGA: AGN Duty Cycle}
\textit{Background:} The time a galaxy spends in the AGN phase, from both general arguments and ensemble studies such as quasar clustering and black-hole mass-function studies and He\,\textsc{ii} proximity-zone analysis, is suggested to last $\sim\!10^{6}$--$10^{9}$ yr. Ionization studies of AGN host galaxies and their surroundings indicate that active nuclei can switch on/off on a $\sim\!100$ kyr timescale -- multiple times during their lifetime -- also finding support in numerical simulations, indicating that the AGN luminosity may fluctuate by orders of magnitude over its lifetime. For a galaxy size of $R\sim10$ kpc it takes $t=R/c\sim30$ kyr for a change in AGN UV luminosity to propagate over the entire galactic disk. We propose to perform a systematic search for fading or, more generally, variable AGN in the most recent MaNGA survey DR17. \\
\textit{Goal:} To assess the total fraction of fading/brightening AGN among the sample, indicative of the duty-cycle fraction these objects span in the transitional phase, directly comparable to the predictions of cosmological simulations. We further aim to characterize the AGN UV luminosity evolution on kyr timescales and compare it, where possible, to that derived from the resolved longitudinal brightness profiles of X-ray jets, potentially yielding a first-of-its-kind link between AGN accretion rate and high-energy particle acceleration in AGN jets.

\subsection{LBG  --  The galaxy--dark matter halo connection of Lyman-break galaxies}
\textit{Background:} A Lyman-break galaxy (LBG) is a galaxy whose broadband photometry shows a ``drop-out'' in the bluest bands as features in its spectrum move from blue to red through the filter set due to cosmic expansion; the reduction in flux blueward of the Lyman-$\alpha$ and Lyman-limit frequencies is caused by scattering and absorption of UV photons by intervening neutral hydrogen. This feature is a robust way to identify galaxies at $z>2$ in photometric surveys, which can then be followed up spectroscopically. Current and future surveys, such as the `Ōnohi`ula Prime Focus Spectrograph Galaxy Evolution Survey (PFS:GE), will measure redshifts for thousands of LBGs at $2.0<z<4.0$ to study their properties and how they trace large-scale structure (LSS). \\
\textit{Goal:} In preparation for the measurement and modeling of the LBG galaxy--galaxy two-point correlation function (2PCF), we will conduct a theoretical investigation into the potential progenitors of the PFS:GE LBG target sample using the Uchuu--UniverseMachine mock galaxy catalog, to better understand how they trace the LSS via galaxy bias and the halo occupation distribution (HOD) as a function of redshift. Of particular interest is any sign of galaxy assembly bias or other selection-function nuances that could lead to incorrect inferences of the galaxy bias ($b_g$) and growth rate of LSS ($f\sigma_8$).
\subsection{IA  --  Intrinsic alignments in varying environments}
\textit{Background:} Weak-lensing surveys are one of the most powerful probes in cosmology; however, the intrinsic alignment (IA) of galaxies contaminates the signal. Many studies have therefore investigated the characteristics of IA in order to eliminate it from the data. So far, researchers believe IA is related to galaxy color and luminosity; in addition, we suspect it is related to the large-scale environment, such as the matter distribution. \\
\textit{Goal:} We will evaluate the additional dependence of IA on the large-scale matter density field and estimate the impact on cosmological analysis when such a dependence is neglected.
\subsection{AR  --  Prediction of debris emergence on laser-ablated sub-wavelength shapes}
\textit{Background:} We have been developing methods to fabricate sub-wavelength structures (SWS) for anti-reflective coating in the millimeter-wave region on hard materials such as ceramics, using ultra-short-pulse laser ablation, which is crucial for machining materials with relatively wide energy band gaps. For efficient fabrication of SWS with a higher ablation rate it is necessary to inject lasers with as high a power as possible; however, at the same time we observe redeposition of ablated debris, particularly at higher energies. Since the physics behind the emergence of debris is mostly unknown, their shape is uncontrollable, and the quantitative conditions under which they appear are ambiguous. \\
\textit{Goal:} To predict the shape expected to be fabricated for a given set of laser-scanning parameters (pulse frequency, pulse energy, spacing of scanning lines, etc.), we first accumulate data on fabricated shapes across different parameter sets and make plots of the results including shape information such as depth and the existence of debris. Our goal is to extrapolate to arbitrary parameters using Gaussian-process regression (and, where useful, large language models), comparing predicted data points against real laser-machining data.
\subsection{RG  --  Radio Galaxies with HalfDome}
\textit{Background:} At low CMB frequencies (around 100 GHz), high-energy radio galaxies act as bright point-source contaminants to CMB maps. The locations of these galaxies are likely correlated with features in the underlying large-scale structure as well as with galaxy properties (e.g., the CIB, radio continuum, X-ray). \\
\textit{Goal:} Add realistic radio luminosities to galaxies in N-body simulations that are relevant for CMB foreground-contamination modeling and correlated with the mass of the underlying haloes, incorporating as much accurate physics as possible (such as the synchrotron-spectrum behavior of different radio-galaxy types). The HalfDome simulations aim to create realistic correlated foregrounds across different wavelengths on N-body simulations.
\subsection{GW  --  Environment of gravitational-wave black hole binaries with weak-lensing maps}
\textit{Background:} The first detection of a gravitational wave (GW) by LIGO opened a new era of multi-messenger astronomy, and around 300 GW events from binary black hole (BBH) mergers have now been observed. However, the origins of these binary black holes and the environments they reside in remain unknown. Combining gravitational waves with cosmological data will provide rich information on the origin and evolution of BBHs. \\
\textit{Goal:} This project aims to unveil the environment of BBHs. We combine publicly available gravitational-wave localization data with cosmological data, such as weak-lensing maps and galaxy-overdensity maps, to investigate the correlation between the BBH distribution and the matter distribution in the universe. The results will be used to constrain the astrophysical origins of BBHs.
\subsection{PTA  --  Forecasting pulsar timing array sensitivity to deviations from general relativity}
\textit{Background:} The Pulsar Timing Array (PTA) is a measurement method relying on the observation of pulsars, fast-rotating neutron stars with well-known timing models. By measuring slight perturbations in the times of arrival (ToAs) of each pulse, computing residuals, and plotting the pulsar-pair correlations as a function of angular separation, one obtains the overlap reduction function (ORF), known in GR as the Hellings--Downs curve. This method allowed, in 2023, evidence for a stochastic gravitational-wave background to be reported by several collaborations (EPTA, NANOGrav, CPTA, \ldots). In modified gravity, additional polarization modes and a modified dispersion relation lead to a modified ORF. \\
\textit{Goal:} In our work we neglect the scalar and vector polarization modes and focus on a modified dispersion relation (which modifies the phase and group velocities). Approximating the gravitational-wave background as plane waves interfering with each other, the relevant quantity to study is the phase velocity, which we do not constrain a priori (parameter space $[0,+\infty)$). Our goal is to forecast the measurement precision required to distinguish, at $1,2,3$ to $5\sigma$ confidence, a $20\%$, $10\%$ to $1\%$ deviation from GR (in terms of phase-velocity deviation), and in how many years that precision will be achievable.
\subsection{SU2  --  Massive Yang--Mills theory}
\textit{Background:} When exploring mechanisms that drive inflation, non-Abelian gauge fields -- such as SU(2) Yang--Mills fields -- have been proposed as alternatives to scalar fields. For instance, introducing a Chern--Simons coupling between a pseudo-scalar field and a non-Abelian gauge field can lead to slow-roll inflation, as in Chromo-Natural Inflation. Moreover, higher-order gauge-field corrections, such as $(F\tilde{F})^2$ terms, can emerge in effective field theories and play a significant role in early-universe dynamics, potentially enabling inflationary scenarios. \\
\textit{Goal:} The goal of this project is to investigate mechanisms by which vector fields can drive cosmic inflation. We focus on Proca and Yang--Mills SU(2) theories and study how to modify them so that their background solution leads to accelerated expansion on cosmological scales, with particular emphasis on the effect of adding a mass term in both theories compared to the massless case.
\section{Literature Review Standardized Prompt}
\label{app:lit_rev_prompt}
Help me find the relevant papers for my project [insert your project ID], which I plan to write a paper about later.
The background of the project is: [insert project background as provided by the human]
The goal is: [insert project goal as provided by the human]
Search for relevant papers that cover the 4 categories:
1. Review papers on the topic (past 15 years)
2. Historically important and highly cited papers on this topic (all time)
3. Recent papers/results on this topic (past 10 years)
4. Other papers that may be relevant
The total number of papers you find (all 4 categories combined) should be no more than 50 papers. Limit to refereed papers only. As for the number of papers in each individual category it is up to you to determine what is appropriate. I want you to for each of the 4 categories create a table of the papers you have found. The columns of the table should be: Title, Author, Journal, Year, DOI, Link.
For "Title" write the full title of the paper.
For “Author” write only the last name of the first author of the paper.
“Year” is the publication year.
For "Link": prioritize ADS, INSPIRE, arxiv links. Other links are OK if those are unavailable.
Do not leave any cells blank. If information is unavailable write NA.
\end{document}